# Bio-inspired Acousto-magnetic Microswarm Robots with Upstream Motility

Daniel Ahmed[1]*, David Hauri[1], Alexander Sukhov[2], Dubon Rodrigue[1], Maranta Gian[1], Jens Harting[2], Bradley Nelson[1]

[1]Institute of Robotics and Intelligent Systems (IRIS), ETH Zurich, Zurich, CH-8092, Switzerland

[2] Helmholtz Institute Erlangen-Nürnberg for Renewable Energy (IEK-11), Forschungszentrum Jülich, 90429 Nürnberg, Germany

*Correspondence and requests for materials should be addressed to D.A. (e-mail: dahmed@ethz.ch, bnelson@ethz.ch)

**Abstract**

The ability to propel against flows, i.e., to perform positive rheotaxis, can provide exciting opportunities for applications in targeted therapeutics and non-invasive surgery. To date, no biocompatible technologies exist for navigating microparticles upstream when they are in a background fluid flow. Inspired by many naturally-occurring microswimmers such as bacteria, spermatozoa, and plankton that utilize the non-slip boundary conditions of the wall to exhibit upstream propulsion, here, we report on the design and characterization of self-assembled microswarms that can execute upstream motility in a combination of external acoustic and magnetic fields. Both acoustic and magnetic fields are safe to humans, non-invasive, can penetrate deeply into the human body, and are well-developed in clinical settings. The combination of both fields can overcome the limitations encountered by single actuation methods. The design criteria of the acoustically-induced reaction force of the microswarms, which is needed to perform rolling-type motion, are discussed. We show quantitative agreement between experimental data and our model that captures the rolling behaviour. The upstream capability provides a design strategy for delivering small drug molecules to hard-to-reach sites and represents a fundamental step toward the realization of micro- and nanosystem-navigation against the blood flow.

**Introduction**

The development of a system with an upstream capability could become a powerful tool that would allow specialized tasks in medicine to be performed, including non-invasive surgical procedures and delivery of drug molecules to hard-to-reach sites. Manipulation of microparticles in a liquid medium can be broadly classified into two approaches: (i) use of external fields or their gradient such as optical(*1*), electrical(*2*) magnetic(*3–7*), thermal (*8*), and acoustic(*9–12*) and (ii) adoption of the Purcell's scallop formulation that require the breaking of the time-reversal symmetry as a design paradigm for motion(*13, 14*). Existing micro- and nanorobots show excellent promise in a biological or biomedical context, however, limited work has been done *in vivo*(*15*). This can be attributed to microscale agents that – when imposed on blood flow such as in a vessel or a capillary – strictly follow the stream with little to no control. To date, manipulation of micro/nanoparticles capable of executing synthetic rheotaxis or upstream movement has been a fundamental challenge. Researchers have evaluated the challenges associated with using strong magnetic fields and field gradients to move microparticles against a flow. However, transport is theoretically limited since the magnetic field gradient scales with its volume and is too low to prompt useful motion. Moreover, the propulsive forces generated from existing micro/nanorobots are insufficient to move under physiological conditions.

Interestingly, many naturally-occurring microswimmers have developed rheotaxis strategies that exploit the non-slip boundary conditions at the wall. Recent studies have shown that the motility of bacteria(*16–18*), spermatozoa(*19–21*), and plankton(*22*) are distinctly different when approaching a boundary. The velocity gradient or shear at the wall generates a torque on cells that in turn generates a wide-range of interesting motion patterns, which are essential for their survival. For example, to collect nutrients from the ocean bed, non-motile phytoplankton, when imposed upon a flow near a boundary, undergo gyrotactic trapping and periodic rotation relative to the direction of the flow. Additionally, when bacteria and spermatozoa become trapped at a boundary, they switch their orientation within the gradient of a shear flow and exhibit rheotaxis, a behaviour commonly used to swim away from predators, collect nutrients, and facilitate navigation of mammalian sperm up the oviduct to assist in fertilization(*23, 24*), as shown in **Fig. 1a**.

In an artificial system, passive or inactivated microparticles cannot physically exploit the boundary condition of a wall when imposed on an external fluid flow. Particles will experience wall-induced lift forces with an inherent direction that acts opposite to the wall, thus preventing them from reaching the wall. Particles need to be energized either chemically (i.e., taking energy from the surrounding environment) or activated by an external acoustic, electric, magnetic or optical field to migrate toward a wall to exploit its non-slip boundary conditions. Inspired by the strategies of microswimmers that exist in nature, researchers have investigated the motion of synthetic swimmers or microrobots near boundaries. The motility of these micro/nanorobots, when confined to a boundary, exhibits remarkable counterintuitive behaviour and, to date, most of these artificial systems are strictly confined to chemical-based or catalytic motors. For example, Simmchen et al. demonstrated docking and controlled navigation of a chemically active catalytic motor on nanometer-sized topological features*(25)* and recently, Das et al. demonstrated navigation of a colloidal motor near a boundary through active quenching of its Brownian rotation combined with hydrodynamic effects near a wall*(17)*. Liu et al. showed that a phoretic swimmer speeds up by a factor of five when restricted to a confined space and this increase in speed is attributed to an electrostatic force combined with electro-hydrodynamic boundary effects*(18)*. Katuri et al. discovered that a catalytic silica-platinum Janus motor exhibited cross-stream migration when imposed on a flow of a hydrogen peroxide solution*(19, 20)*. Palacci et al. demonstrated that a realigned biomimetic catalytic motor within the shear flow gradient of hydrogen peroxide solution exhibited rheotaxis when exposed to blue light*(20, 21)*. Recently, a bimetallic micromotor was shown to undergo rheotaxis in a combination of a chemical fuel and acoustic force*(31)*. Even though chemically driven microrobots exhibit rich dynamic motion and numerous fundamentally new behaviours near a boundary, their application in medicine is limited, especially in anatomical locations such as the vasculature system where robots need to be controlled within an imposed blood flow. However, previous work showed that chemically-driven microrobots could function as an effective drug delivery platform for treating infection in the stomach of mice*(23)*. Current synthetic systems capable of cross- or upstream movement require a foreign chemical fuel and the upstream velocities attained are only a few body lengths per second within a low-imposed stream of $10 \, \mu m/s$. To date, no micro/nanorobotic system has been described that can perform rheotaxis in an external field. The use of externally-applied acoustic or magnetic fields to induce rheotaxis would be extremely advantageous *in vivo*, since toxic chemicals would not be needed and this type of approach would be regarded as non-invasive and biocompatible.

The development of upstream migration could become an important transport mechanism in physiological conditions as most other approaches would fail due to the high flow rates and the lift-forces resulting from interaction between the particles and the wall. In this article, we present a mechanism that allows swarms of microparticles to be transported against an imposed flow along the boundaries of a microchannel using a combination of externally actuated acoustic and magnetic fields. Each microswarm is formed by the self-assembly of individual superparamagnetic particles resulting from dipole-dipole interactions in a rotating magnetic field. The spinning microswarm is then acoustically directed toward the wall where it executes rolling and subsequent rheotaxis. We estimated the magnitude of the acoustically-induced reaction force of the microswarm from the wall, which is essential for rolling. We characterized rheotaxis of a microswarm at different imposed flow rates and observed quantitative agreement between experimental data and the rolling model. Microswarm rheotaxis was quantitatively studied and upstream motion against the stream (as high as $1.2 \text{ mm/s}$) was observed, after which the swarms lost their traction and eventually were dragged away by the flow. The presented upstream capability provides a potential design strategy for delivering drugs to hard-to-reach capillaries and represents an important step toward the realization of micro/nanosystems navigation against the blood flow.

## Results
### Concept of Synthetic Rheotaxis

Inspired by rheotaxis of bacteria and spermatozoa that exploit the non-slip boundary conditions of the wall, we devised a new strategy for guiding swarms of micro/nanoparticles against a stream using a combined acoustic and magnetic field, see **Fig. 1a**. Researchers have shown manipulation of colloidal microrobots near walls using magnetic fields(*33*, *34*), and in a combination of acoustic and magnetic actuation particularly in quiescent fluids(*35*). Navigation of particles against a background flow becomes a fundamental limitation since most particle manipulation systems, will fail when exposed to an external flow. The forces that are generated by most micro and nanorobots are sufficiently insignificant to propel against background fluid flows. Furthermore, the hydrodynamic interactions of moving microrobots within the channel wall give rise to wall-induced lift forces, which prevent particles from reaching the wall to exploit the boundary conditions of a wall. Here, we investigate the upstream propulsion of microswarms in the presence of a background flow. The onset of motion in a fluid flow requires a reaction force from the wall that is compensated for by an acoustic radiation force exerted on the microswarm. The acoustic-induced reaction force further assists in overcoming the oppositely directed wall-induced lift force, enabling the microswarm to reach the wall and to execute rolling. An essential aspect of our system is that it exploits the flow characteristics near a wall. Due to the high shear force and low velocity at a wall that result from the wall's non-slip boundary conditions, manipulation of particles becomes favourable for motion against a flow. The concept of upstream behaviour is schematically shown in **Fig. 1b**. The superparamagnetic microparticles assemble into a spinning microswarm in an externally imposed rotating magnetic field $\boldsymbol{B}(t) = B_0 \cos\omega t \boldsymbol{e}_x + B_0 \sin\omega t \boldsymbol{e}_z$. In the presence of an acoustic field, the microswarm is pushed vertically toward the wall and a region of minimum pressure to execute upstream rolling.

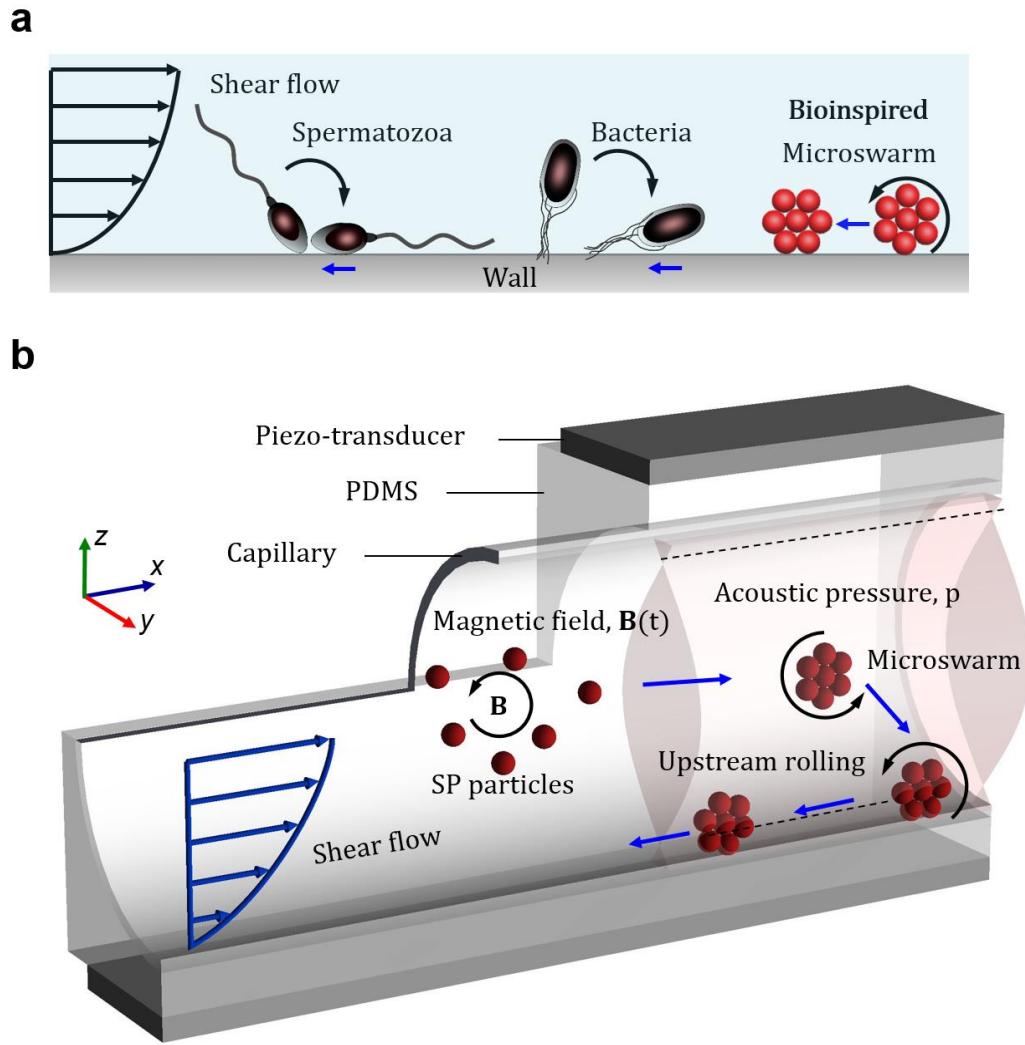

**Fig. 1 | Bio-inspired upstream motion in an acoustic and magnetic field.** (**a**) Naturally-occurring microswimmers such as spermatozoa and bacteria exploit the non-slip boundary conditions to execute upstream motion along the wall. (**b**) In a bioinspired system, injected superparamagnetic (SP) particles in the capillary self-assemble into a spinning microswarm due to the dipole-dipole interaction of a rotating magnetic field. The microswarm was shown to be marginalized toward the wall in an acoustic pressure field. The microswarm executes upstream rolling and performs subsequent rheotaxis when imposed on external liquid flow.

**Microswarm Manipulation in an Acoustic Field**

Acoustic waves or high-frequency ultrasound are particularly attractive because of their ability to penetrate deep into the tissue. They are not affected by non-transparent or opaque bodies of animal models, and can produce a steady force on an object. Here, we discuss the acoustic manipulation of microparticles by introducing a couple of arrangements of the acoustic setups, which enable us to estimate the acoustic-induced reaction force. This reaction force, which acts on microswarms, becomes critical to perform rolling and subsequent rheotaxis.

The working mechanism of our custom-built acoustofluidic device is shown in **Fig. 2**. The device is comprised of a 3D printed chassis of dimensions $10 \times 2 \times 1 \text{ mm}^3$, piezo-transducers, and a circular glass capillary embedded in a polydimethylsiloxane (PDMS)-polymer (see **SI Text**). Circular microchannels of diameter $150$ and $300 \text{ μm}$, were implemented to mimic physiological flow conditions. A computer-controlled programmable syringe pump was used to regulate liquid flow within the acoustofluidic device. An electronic function generator controlled the excitation frequency and the power applied to the piezo-transducers. The microparticles in the liquid medium were acoustically manipulated using an excitation frequency of $2.0 - 7.2 \text{ MHz}$ at $20 \text{ V}_{\text{PP}}$. The vibration of a piezo-transducer generates bulk acoustic or pressure waves in a liquid medium. When the oppositely-paired piezo elements were actuated at their thickness mode, the propagating waves created a 1D longitudinal standing pressure wavefield across the channel (3). The interference of two acoustic wavefield series produces a periodic distribution of minimum (node) and maximum (antinode) pressure locations. The pressure gradient exerts a force on the suspended spherical microparticles, which generally pushes them toward the nodes. Most micro-objects, such as polystyrene beads, superparamagnetic particles, and living cells, become trapped at the pressure nodes (*11*, *36*). **Fig. 2a** and **b** shows the rapid arrangement (within ~1 seconds) and the fluorescent microparticles trapped between adjacent pressure nodes upon excitation of the acoustic field (see also **Movie S1**).

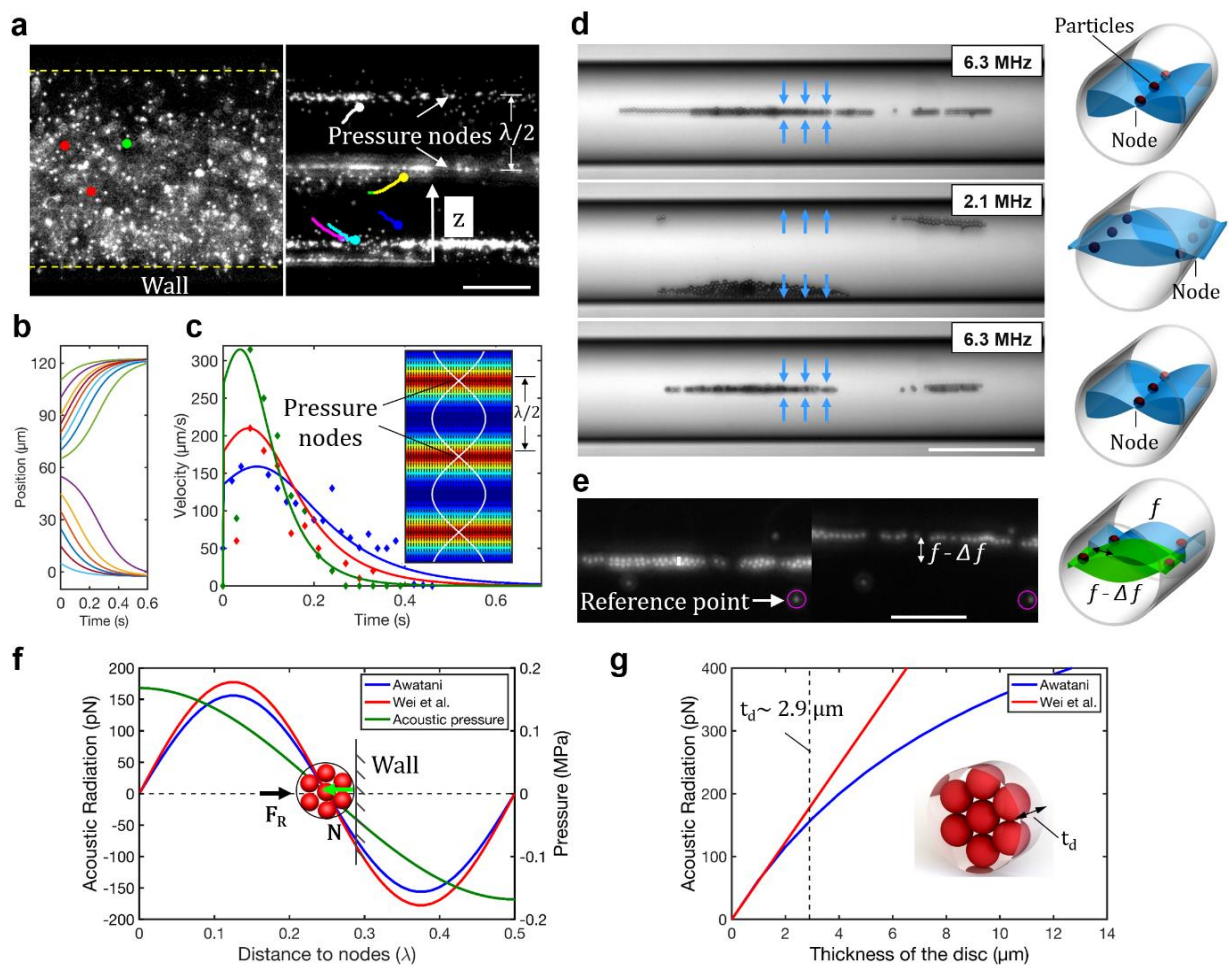

**Fig. 2 | Estimation of the acoustic-induced reaction force.** (**a**) Randomly distributed fluorescent particles are organized in the pressure nodes of a standing acoustic wavefield of a 300 μm diameter capillary. The trajectories of individual particles toward the nodes are shown in white, magenta, cyan, blue, and yellow colors (see also **Movie S1**). Scale bar 150 μm. (**b**) Position versus time of microparticles migrating toward the nodes. (**c**) Velocity versus time of tracked microparticles moving toward the nodes. The inset shows the simulated acoustic pressure field between nodes. (**d**) Microparticles are manipulated from the center to the sidewall and vice versa of a 150 μm diameter capillary by switching the resonant modes of the piezo transducers (see **Movie S2**). The blue and green color in the cartoon schematic represents the standing wavefield developed within the capillary. Individual particles are trapped in the pressure nodes. Scale bar 150 μm. (**e**) A linear band of trapped fluorescent microparticles in a node was shifted spatially by manually sweeping the excitation frequency (see **Movie S3**). A white dot, circled in magenta, indicates the static reference point. Scale bar 50 μm. (**f**) We considered a couple of analytical solutions of the radiation force: when the direction of the standing wavefield is (i) perpendicular (Awatani et al., see also **SI Text, Note S2**) and (ii) parallel (Wei et al.) to the surface of the disc to estimate acoustically induced reaction force. Although the two models have small discrepancies in the force estimation, the model extracted by Wei et al. is experimentally more relevant, since the direction of the propagating waves is parallel to the surface of the disc-shaped microswarm. (**g**) The plot demonstrates the acoustic radiation forces on self-assembled microswarms versus their thickness.

As acoustic waves propagate through a liquid containing microparticles, they impart acoustic radiation forces on the particles. The radiation force that acts on the particles is comprised of primary and secondary radiation forces. The primary radiation force results from the interactions of the scattered waves from the particles within the background standing acoustic waves, while the secondary radiation force results from the interparticle interaction of the scattered waves from the neighbouring particles. For simplicity, we ignored the secondary interaction of the particles in an acoustic field. The primary radiation force, $F_R$, acting on a single isolated incompressible particle in an acoustic pressure field can be estimated by the gradient of the Gor'kov potential. The time-averaged force acting on a particle of radius $R$, where $R$ is significantly smaller than the wavelength $\lambda$, in a one-dimensional standing wave of wavenumber $k_z = 2\pi/\lambda$ and pressure field $p = p_{ac} \cos(k_z z)\cos(\omega t)$, can be expressed analytically as (24)

$$\boldsymbol{F}_R = 4\pi\phi(\tilde{\kappa}, \tilde{\rho}) k_z R^3 E_{ac} \sin(2k_z z), \qquad \text{eq. (1)}$$

where $\phi(\tilde{\kappa}, \tilde{\rho}) = \frac{1}{3}\left[\frac{5\rho_s - 2\rho_0}{2\rho_s + \rho_0} - \frac{\kappa_s}{\kappa_0}\right]$ is the acoustophoretic contrast factor; $\rho_0$ and $\rho_s$ denote the density of the liquid and particle, respectively; $\kappa_0$ and $\kappa_s$ denote the compressibility of the water and particle, respectively; $p_{ac}$ denotes the acoustic pressure amplitude; $E_{ac}$ denotes the acoustic energy density; and $z$ determines the position of the particle within the field. The acoustic energy density scales as $E_{ac} = KV_{PP}^2$, where $K$ is a relation constant that permits the determination of the acoustic radiation force experimentally and $V_{PP}$ is the voltage applied to the piezo transducers. The acoustic contrast factor $\Phi$ determines whether the microswarm moves toward the node or the antinodes of the wavefield. $\Phi$ is positive for superparamagnetic particles and is measured to be 0.3, suggesting that the particles will move to the nodal lines, see also **SI Text, Note S1**.

We utilized the Gor'kov's potential of a single microparticle trap to measure the acoustic energy density $E_{ac}$ and the applied pressure $p_{ac}$, which we later used to estimate the acoustic-induced reaction force. The acoustic radiation force acting on a single superparamagnetic particle is balanced by the Stokes drag, $F_D$, measured as $\boldsymbol{F}_D = \boldsymbol{F}_R = 6\pi a\eta \boldsymbol{v}_z(t)$, where $\boldsymbol{v}_z$ is the velocity of the particle approaching the pressure nodes. Knowing the expression of the acoustic radiation force, the velocity function of the particle can be arranged as $\boldsymbol{v}_z = \frac{F_R}{6\pi\eta R} = \frac{2\phi k_z R^2 E_{ac}}{3\eta}\sin(2k_z z(t))\boldsymbol{e}_z$, where $z(t)$ is the displacement of the particle within adjacent pressure nodes. The trajectories of the microparticles reaching the pressure nodes were tracked manually in ImageJ. The velocity function is fitted to the experimental data in **Fig. 2c** to estimate $K$ and the corresponding $E_{ac}$. We then measured the acoustic pressure applied to the acoustofluidic device as $E_{ac} = p_{ac}^2/4\rho_o c_o^2 \sim 1.1 - 3.2 \text{ J/m}^3$; hence, $p_{ac} \sim 0.098 - 0.2 \text{ MPa}$. These measured values are of the same order of magnitude, as discussed in the literature(*47, 48*). The small discrepancies can be attributed to the various designs of acoustofluidic devices.

We adopted two methods to acoustically manipulate spherical microparticles toward the wall to facilitate the onset of rolling and initiation of rheotaxis. In the first method, the superparamagnetic particles in a water solution are driven toward the wall by modulating the resonant modes of the piezoelectric transducers. **Fig. 2d** shows that microparticles become trapped at the center of a capillary (150 μm) when the piezo-transducers are actuated at their second resonant mode (6.3 MHz). The particles are then shifted toward the wall nearly instantaneously by switching the transducers to their first mode (2.1 MHz), see also **Movie S2**. The distance between adjacent nodes is half the acoustic wavelength $\lambda = c/f = 350$ μm, where $c = 1498 \text{ ms}^{-1}$ is the speed of sound in water at 298 Kelvin and $f = 2.1$ MHz is the applied frequency of the ultrasound. Since the distance between adjacent nodes is larger than the diameter of the capillary, at least one nodal position must be located outside the capillary. As a result, microparticles are pushed toward and become trapped against a wall.

In the second method, the acoustic trap, i.e., the pressure node was designed to appear near a wall. A spinning microswarm does not always initiate rolling when approaching a wall, which can be attributed to the negligible acoustic-induced reaction force. We compensated for the lack of reaction force by manually adjusting the excitation frequency such that the microswarm was spatially shifted toward the wall until rolling was observed. Most commercially available piezoelectric transducers have a short operating frequency or low bandwidth; i.e., the quality factor is reasonably sharp. However, a limited shift in the excitation frequency is still sufficient to produce a noticeable spatial change in the position of the trapped particle. **Fig. 2e** shows trapped microparticles in a pressure node arranged in a line, which is spatially shifted by sweeping the acoustic frequency, see also **Movie S3**.

**Estimation of the Acoustic-induced Reaction Force**

We next estimated the magnitude of the acoustic-induced reaction force of the microswarm from the wall, which becomes important for rolling. The assembled microswarm was experimentally found to be disc-shaped with thickness $t_d$; as a result, the microswarm's geometry into the force measurement had to be considered. Wei et al. developed an analytical solution of a radiation force on a thin, rigid disc of radius $R \ll \lambda$ in a 1D standing wavefield, in which the incident acoustic wave is parallel to the surface of the disc(*38, 39*)

$$F_R = \frac{R}{4} p_{ac}^2 \kappa_o t_d \beta_{st} \sin(2k_z z), \qquad \text{eq. (2)}$$

$$\beta_{st} = \pi k_z R \left[\left(1 - \frac{\kappa_s}{\kappa_o}\right) + 2\left(\frac{\rho_s - \rho_o}{\rho_s + \rho_o}\right)\right], \quad \text{eq. (3)} \qquad \kappa_s = \frac{3(1-2\nu)}{E}, \quad \kappa_o = \frac{1}{\rho_o c_o^2}, \qquad \text{eq. (4)}$$

where $E \sim 30\ GPa$ denotes the Young's modulus and $\nu = 0.33$ denotes the Poisson ratio of the superparamagnetic microparticle. We assume the microswarm to be a rigid elastic body, which satisfies $R/\lambda \ll 1$. **Fig. 2f** shows the sinusoidal dependence of the analytical solution of the radiation force acting on the microswarm. When the microswarm is at the pressure antinode, the maximum radiation force is achieved; i.e., at $\lambda/4$ from the nodal position.

As the microswarm comes in contact with the wall, the acoustic radiation force becomes the normal reaction force, $N$. The radiation force at the center of a microswarm of $R = 15$ μm is zero because it is trapped at the pressure node, i.e., $p_{ac} = 0$. However, at $\pm 15$ μm from the center of the swarm, a non-zero force exists, which produces the acoustic-induced reaction force. The minimum reaction force needed to execute rolling is $\sim 15\%$ of the maximum radiation force $|F_R|_{max}$ in Wei's Model in **Fig. 2g**; then $N \approx 25$ pN. For a microswarm of $R \sim 20$ μm, $N \approx 30$ pN is needed to execute rolling. Thus, the minimum acoustic-induced reaction force needed to execute rolling can be estimated as $N \geq \frac{R|F_R|_{max}}{1 \times 10^{-4}}$. The acoustically induced reaction force, which depends on the incident acoustic pressure, the surrounding liquid environment, and the intrinsic property and geometry of the microparticle (**Fig. 2f**), can be tailored for effective rolling.

**Self-assembly of Microswarms in a Rotating Magnetic Field**

In this section, we study the formation of microswarms in combined acoustic and magnetic fields. The frequency and magnitude of the external oscillating magnetic field determine the formation of superparamagnetic microswarms. A magnetic field $B_0 = 10$ mT ensures that the dipole–dipole interactions are large enough to attract neighboring particles. The field strength, $B_0$, is of the order of the coercive field of single particles, i.e., $B_c \approx [1 - 10]$ mT at room temperature. (See **SI Text, Note S3**.) This might lead to two scenarios, i.e., i) the magnetization can jump over the potential barrier induced by the magnetocrystalline anisotropy, leading to no net particle rotation around its own axis ($B_0 > B_c$), and ii) the microparticles can perform rotations caused by the external field ($B_0 < B_c$). Our preliminary measurements suggest the second scenario.

We estimate the Reynolds number of the microswarm to be low, i.e., $Re \approx 4 \cdot 10^{-4}$, which ensures the dominance of the viscous over the inertial forces and the overdamped type of particle motion. Hence, to understand the transition where linear chains of single particles are transformed into disc-shaped circular microswarms, we consider a dimensionless entity known as the Mason number $Mn$ (*40*, *41*). We employ an expression of Mn, which is the ratio of viscous to magnetic forces, for a chain of $N$ magnetic beads as $Mn = 16 \frac{\eta\omega}{\mu_0 \chi_p H_0^2} \frac{N^3}{(N-1)\left[\ln\left(\frac{N}{2}\right)+\frac{2.4}{N}\right]}$, where $\eta$ denotes the dynamical viscosity of water, $\omega$ is the driving frequency of the oscillating magnetic field, $\mu_0$ is the magnetic permeability in vacuum, $\chi_p$ is the particle susceptibility and $H_0$ stands for the strength of the oscillating magnetic field. For $Mn < 1$, the chains produced are stable and can rotate as a whole (*42*), which is the case for low $\omega$ and small $N$. For $Mn > 1$, the chains begin to fragment into microswarms. Based on the equation with parameters $N = 30$ and $B_0 = 10$ mT, we expect the linear chain to fragment at frequencies of several Hz.

When superparamagnetic solutions were introduced in the quiescent fluid within a capillary, microparticles were pushed toward the nodal lines and became trapped in the standing acoustic wavefield, as shown in **Fig. 3a**. When an external rotating magnetic field of $10\ mT$ at $22\ Hz$ was applied, spinning microswarms were produced almost instantaneously in these nodal lines, as illustrated in **Fig. 3b** and **c** (see also **Movie S4**). We further investigated the formation of microswarms when exposed to external fluid flow. Since the Reynolds number of the superparamagnetic microparticles is significantly lower than 1, viscous forces dominate. Single microparticles would not drift due to inertia. Microparticles remained in close proximity relative to each other in a bulk flow in the absence of acoustic and magnetic fields. When a standing acoustic field was applied, the superparamagnetic particles became trapped and concentrated within the minimum pressure bands. When an external rotating magnetic field was applied, the neighbouring superparamagnetic particles produced microswarms within the acoustic bands in an assembly-line fashion, see also **Movie S4** (bottom section).

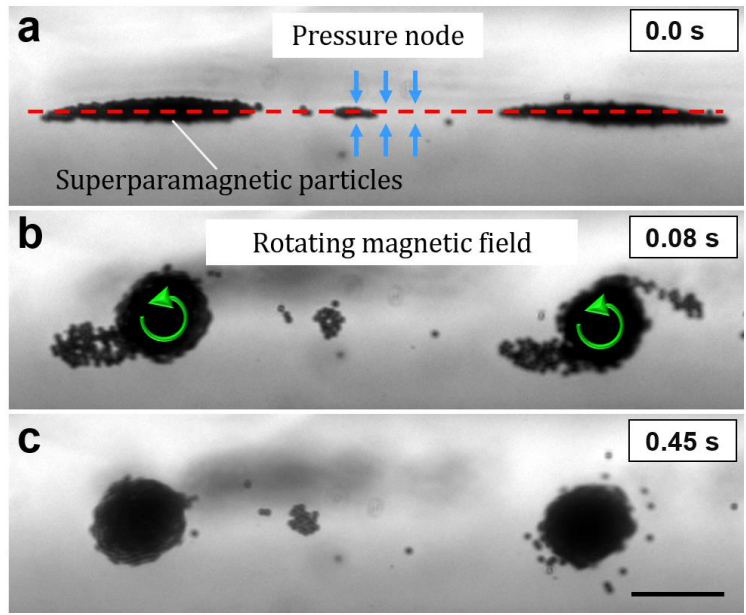

**Fig. 3 | Microswarm formation.** Image sequences demonstrate the formation of the microswarms in an externally imposed rotating magnetic field within an acoustic pressure node. The superparamagnetic particles were arranged in the nodal line of a standing acoustic wavefield. An applied rotating magnetic field transforms the trapped superparamagnetic microparticles into rotating microswarms (see also **Movie S4**). The scale bar corresponds to 45 μm.

**Onset of Upstream Rolling upon an Imposed Flow**

Here we present a theoretical model of rheotaxis of the microswarm rolling along the walls of the microchannel upon an imposed flow. Each self-assembled microswarm is regarded as a rigid body having a disc form with radius $R$ and thickness $t_d$ (**Fig. 4**). Observations show that microswarms spin much faster than they roll, especially in the vicinity of the wall. This suggests a fluid layer between the spinning microswarm and the wall which requires an approach based on a wet wall (lubrication effect) rather than a dry wall (non-slip boundary conditions). The microswarm experiences multiple forces and torques as it reaches the boundary, as schematically shown in **Fig. 4a**. Since the Reynolds number of the microswarm is low, both translational and rotational types of motion are overdamped, meaning that the sum of all forces and torques is zero $\sum_i \boldsymbol{F}_i = 0, \sum_i \boldsymbol{T}_i = 0$.

Along the propagation direction of the microswarm, the friction force $F_F$ is balanced by the Stokes' drag force $F_D$, where $F_F = \mu_{WF} \cdot N$ (**Fig. 4**). Here, the wet friction coefficient is $\mu_{WF}$ and $N$ is the contact force, which is induced acoustically. The modification of the Stokes drag force should take into account both the disc-like shape of the microswarm and its proximity to the wall. Therefore, we denote the effective Stokes drag coefficient by $k_D$ and consider it further as a fitting parameter. At low Reynolds number, the time-reversal nature of the spinning microswarm ensures no motion. To generate noticeable movement, the microswarm needs to break either the symmetry of its rotational motion or the symmetry of its surrounding geometry. The resulting horizontal force balance equation reads $F_F = F_D$ or $\mu_{WF} \cdot N = k_D \cdot \eta \cdot D \cdot \left[V_1 + V_2|_{\left(\frac{D}{2}+h\right)}\right]$, where $V_1$ is the microswarm velocity, $V_2$ is the fluid velocity at the microswarm center of mass, $D$ is the diameter of the microswarm, and $h$ is the gap between the wall and the microswarm. The gap length was experimentally estimated as $h = 1.8 \pm 0.3$ μm, estimated from at least 10 measurements, as shown in **Fig. 4b** and **c**. We disregarded the effect of lift forces, which together with the gravitation force should be balanced by the acoustic forces (*45*). Also, the effect of Brownian motion was not included in the model, since the ratio of thermal to viscous forces is negligible for a microswarm of $D/2 \approx 30\ \mu m$ moving at $V_1 \approx 100\ \mu m/s$ (see **Materials and Methods**).

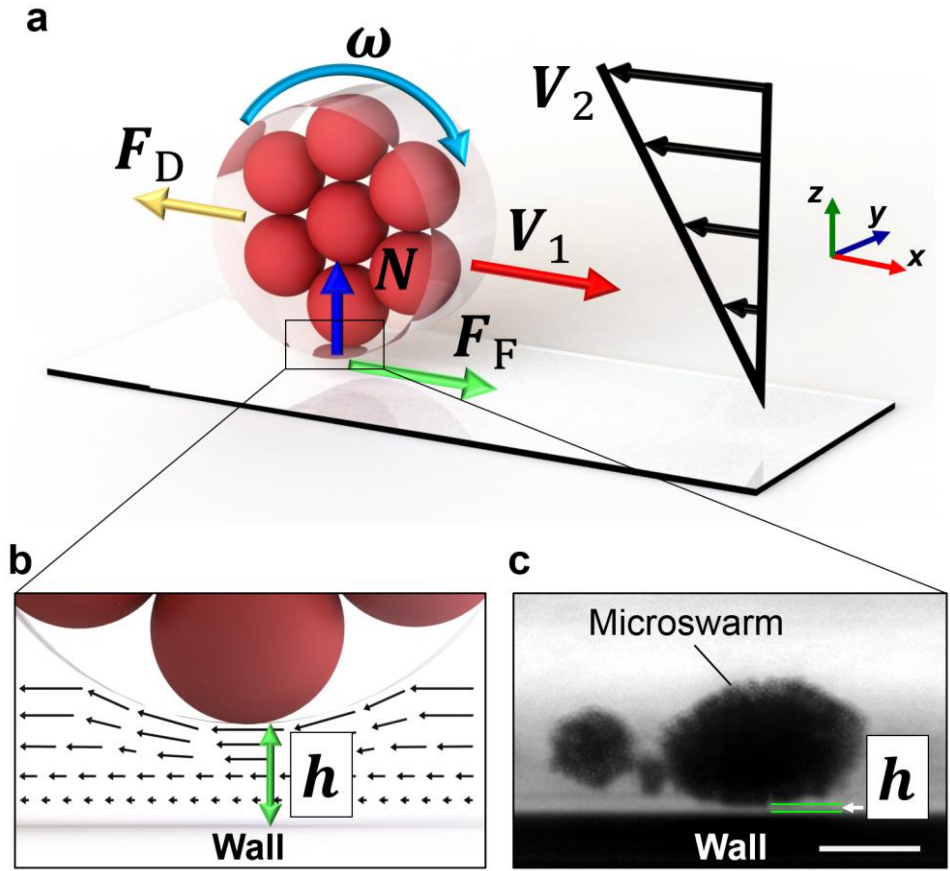

**Fig. 4 | Upstream rolling model of the microswarm.** (**a**) The schematic shows the forces and torques acting on a rolling microswarm when imposed by the external liquid flow. (**b**) The inset represents the torque produced by the lubrication layer between the spinning cluster and the wall. (**c**) Experimental estimation of the gap length, $h$ between the wall and the microswarm.

In addition to the multiple forces, the microswarm is strongly subjected to a number of torques. The torque due to the friction force $\boldsymbol{T}_F = \mu_{WF} \cdot \boldsymbol{N} \cdot \frac{D}{2}$ is the difference between shear torques at the bottom $\boldsymbol{T}_{Sh}(h)$ and the top $\boldsymbol{T}_{Sh}(D+h)$ of the microswarm, i.e., $\boldsymbol{T}_F = \boldsymbol{T}_{Sh}(h) + \boldsymbol{T}_{Sh}(D+h)$ or $\mu_{WF} \cdot \boldsymbol{N} \cdot \frac{D}{2} = \eta \cdot \left(\frac{\omega D}{2} - V_1 + V_2|_{(h)}\right) \cdot A \cdot \frac{D}{2h} + \eta \cdot \left(-\frac{\omega D}{2} - V_1 + V_2|_{(D+h)}\right) \cdot A \cdot \frac{D}{2(D+h)}$. Here $A = Dt_d$ denotes the stressed area of the microswarm. From the forces and the torques, we derived a relation of the microswarm upstream velocity, which strongly depends on its rotational frequency, radius, and the imposed flow velocity:

$$V_1 = c_1 \omega + c_2 V_2|_{(h)} - c_3 V_2|_{\left(\frac{D}{2}+h\right)} - c_4 V_2|_{(D+h)}, \quad \text{eq. (5)}$$

where coefficients have the following form $c_1 = \frac{A(D+2h)D}{2(k_D hD(D+h)+AD)}$, $c_2 = \frac{A(D+h)}{(k_D hD(D+h)+AD)}$, $c_3 = \frac{k_D h D (D+h)}{(k_D hD(D+h)+AD)}$ and $c_4 = \frac{A h}{(k_D hD(D+h)+AD)}$.

**Motility or Rheotaxis**

We investigated upstream motility or rheotaxis of self-assembled microswarms imposed upon a stream in a combined acoustic and magnetic field. We injected a water solution containing 2.9 μm superparamagnetic particles into our acoustofluidic device. The flow profile within the capillary was quantified by particle-tracking and the data fit well with the Poiseuille flow of a circular channel (see **SI Text, Note S5**). In an externally applied rotating magnetic field, the particles self-assembled into spinning microswarms. The microswarm migrated toward the capillary wall when an acoustic field was applied (see also **Movie S5**). The spinning microswarm broke its rotational symmetry when it reached a wall, which contributed to a rolling-type translational motion. We characterized the upstream rolling of both acoustic manipulation conditions where the pressure node lies 1) outside and 2) inside the capillary.

**Fig. S6** and **Movie S6** illustrate the upstream motion of the superparamagnetic microparticles against a $\sim 50\ \mu m/s$ flow stream in an acoustic arrangement where the pressure nodes reside outside the capillary, as shown in **Fig. 2d**. The superparamagnetic particles were initially trapped acoustically at the center of the capillary. In the presence of a rotating magnetic field, these microparticles assembled into microswarms. When the excitation acoustic frequency was switched from the second to the first resonant mode of the piezo, the spinning microswarms fragment into individual particles when reaching the wall and exhibited an upstream rolling motion from left-to-right. Even though this acoustic setup provided satisfactory upstream motion, single particle analysis and tracking close to the curved capillary were difficult. We devised a different acoustic arrangement such that microswarms did not fragment when they reached the wall.

**Fig. 5a** shows the microswarm's rolling upstream along the wall in an acoustic arrangement in which the pressure node is located inside the capillary. A microswarm that was spinning counter-clockwise exhibited left-to-right rolling when it reached the wall, which is traced with a blue trajectory (also see **Movie S7**). The cyan, green, and red trajectories show that the individual particles that were unaffected by the acoustic wavefield followed the streamlines of the flow. Although both acoustic arrangements effectively produced upstream rolling, the major distinction in this arrangement was that the microswarm did not break up when it reached the wall. The current arrangement enabled each cluster to maintain its nearly circular shape, thus making it easier to track the images. **Fig. 5b** shows the upstream velocity, $V_1$, of the microswarm against the increasing velocity of the stream, $V_2$. We considered the maximum upstream velocity for at least five measurements from multiple spinning clusters for each flow rate. The fastest translating clusters were identified and traced by a tracking plugin in **Image J**. At low-flow velocities (i.e., in the range of $0.0 - 0.2 \text{ mm/s}$), the upstream rolling was affected marginally by an externally-imposed flow. As the flow rate, $V_2$, was increased gradually, the upstream velocity, $V_1$, decreased. Upstream rolling or positive rheotaxis was observed against a stream velocity as high as $1.2 \text{ mm/s}$, after which the microswarms lost their traction from the wall and eventually were swept away by the flow. The speed of the microswarms scales as $-V_2$ in water. **Fig. 5b** shows that this linear relationship (red and blue fits) was satisfied reasonably by the microswarms rolling along the wall. The slight deviation of the linear fit, i.e., their slopes, was expected given that (i) the thickness of the microswarm changes during rheotaxis, which introduces variability in the stressed area, $A$, parameter, (ii) variability in the microswarm's drag coefficient, (46) (iii) rolling microswarms are not entirely circular in shape, and (iv) the relatively large distribution of the microswarms' diameter during rheotaxis. Next, we studied upstream rolling versus the microswarm's diameter, as shown in **Fig. 5c**. The upstream values also were reasonably close to the anticipated prediction of equation (5), as shown in **Fig. 5c** (red fit). **Fig. 5c** (blue fit) demonstrates the upstream rolling of the microswarm in the absence of any flow. Although our simple model does not take into account the deformability of the microswarm (Eq. (5)), it correctly describes the upstream behaviour of the microswarms. Also, the model gives the correct scaling for the increasing radius of the microswarm at fixed imposed flow (**Fig. 5c**). The main result of the work is the demonstration of the microswarm propulsion against the flow, and **Fig. 5** quantifies this finding.

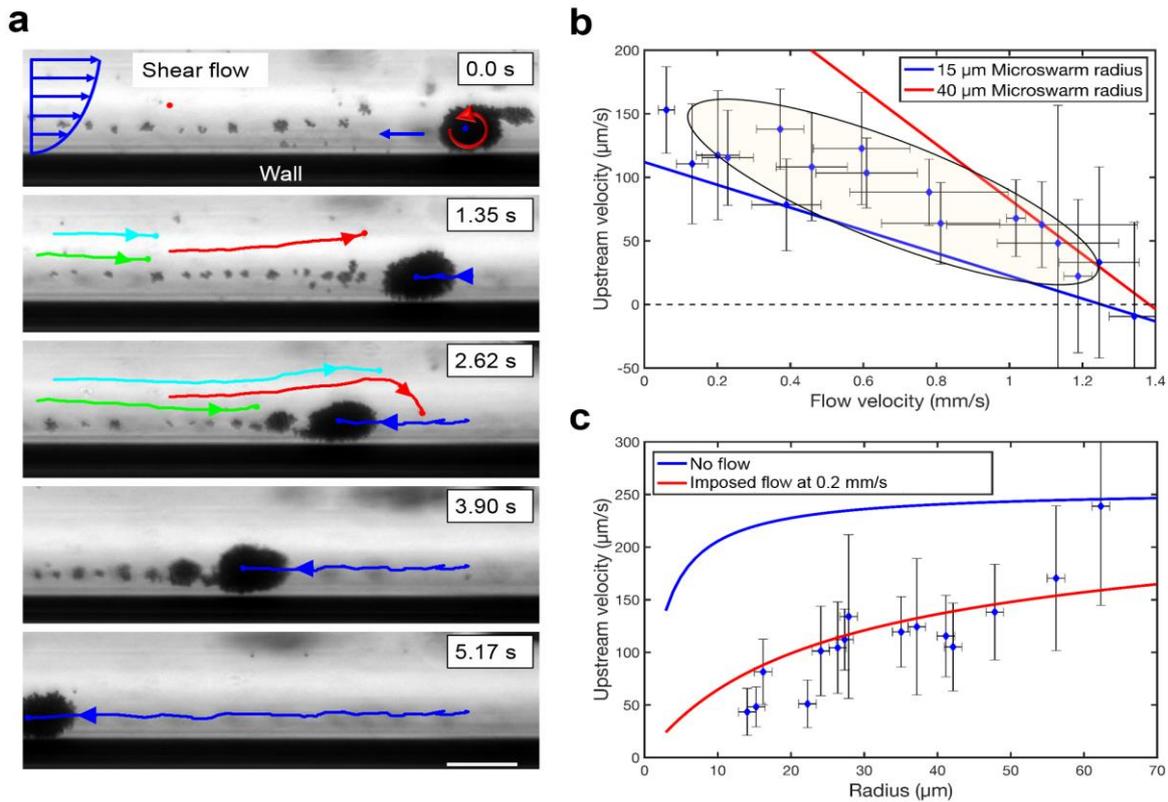

**Fig. 5 | Microswarm rheotaxis in combined acoustic and magnetic fields.** (**a**) Image sequences demonstrate the rheotaxis of a microswarm rolling along the capillary's wall (tracked in blue) in a combined acoustic and magnetic field. The streamlines of the oppositely-directed flow are shown by cyan, green, and red colored trajectories (see also **Movie S7**). The scale bar corresponds to 60 μm. (**b**) The plot demonstrates the microswarms' upstream velocity ($V_1$) versus imposed flow rates ($V_2$) within the fluidic channel. The upstream model was fitted for 15 and 40μm cluster sizes, respectively. All measurements in the plot were performed using an excitation acoustic frequency and voltage of ~6 MHz and 20 $V_{PP}$, respectively, at a rotational magnetic field of magnitude and frequency of 17.5 Hz at 20 mT. (**c**) The plot characterizes the upstream velocity ($V_1$) versus the dimension of the microswarms at an imposed flow rate of 0.2 mm/s. The red fit, which is based on Eq. 5, maps well with the experiment measured values for rolling at $h = 1.8$ μm. The blue-colored fit is mapped on the plot for no flow. Error bars represent the standard deviation (s.d.). Each data point represents the average rolling velocity of 3–5 swimmers.

**Discussion**

This work demonstrates a bio-inspired, wireless micro/nanorobotic system that exploits the non-slip boundary conditions of a wall in a combination of externally-triggered acoustic and magnetic fields. Our system produced, *for the first time*, upstream motility of microswarms using biocompatible fields. External, field-driven microrobotic systems are particularly attractive because they do not require an on-board power supply or intricate moving parts. Thus, they can possibly be scaled down to a few hundreds of nanometres in size. Although, at the nanometre scale, the influence of thermal agitation or Brownian motion should be taken into consideration, our system, i.e., the formation and navigation of microswarms, is robust against thermal noise for the sizes that have been presented (see **Supplementary Note S4**). The self-assembled robotic microsystem has numerous advantages over prefabricated micro/nanorobots. For example, multiple steps and complicated fabrication methods restrain the rapid mass-production of many types of prefabricated micro/nanorobots. In contrast, we used commercially-available, biocompatible, superparamagnetic particles with dimensions of about 3 micrometres. The assembled microswarm, which is produced by the action of magnetic dipole-dipole interactions, is better suited for rolling than its solid counterparts. Under high-shear flow, the solid counterpart may lose traction from the wall, whereas the assembled microswarm deforms its shape drastically to remain in contact with the wall to initiate rolling. **Movie. S8** shows the deformation of the microswarm by more than 30%.

We believe that the technology we have developed eventually can be translated into *animal models.* The blood flow velocity in small animal models, such as rat, mouse, and zebrafish, is in the range of a few $mms^{-1}$. Considering the findings in **Fig. 5**, we believe the proposed method will find its usefulness in animal bodies. Although the inability to image microrobots *in vivo* has been a major drawback in the field of micro/nanorobotics [1], recent studies have shown real-time tracking of microrobots using photoacoustic imaging in tissue phantoms and intestines of mice models [2, 3]. We plan to incorporate our methodology with appropriate imaging modality to visualize and navigate robots *in vivo* in small animals.

**Materials and Methods**

**Acoustofluidic Device and Experimental setup.** The fabrication procedure of the 3D-printed acoustofluidic device was discussed in the **SI Text**. Details of the acoustic and electromagnetic experimental setup was described in the **SI text**. **Data availability.** The authors declare that data supporting the findings of this study are available within the paper and its supplementary information files.

**Acknowledgement.** The work was supported by ETH Zurich Career Seed Grant-14 17-2. This work has been financed by the European Research Council Advanced Grant (SOMBOT), Grant Agreement No. 743217; European Research Council Starting Grant (SONOBOTS), Grant agreement No. 853309; and DFG Priority Programme SPP 1726, Microswimmers – "From Single Particle Motion to Collective Behaviour." We would like to thank Ayoung Hong, Nicolas Blondel, and Nicolas Sieber for their helpful discussion. **Author contributions.** D.A. initiated and designed the project. D.A contributed to the experimental design and scientific presentation. D.H., M.G., A.S., and D.A. performed all the experiments and data analysis. A.S., D.R., D.A., and J.H. developed the theoretical studies. D.A. wrote the manuscript with contribution from all authors. All authors contributed to the scientific discussion. **Competing interests**. The authors declare no competing financial interests.